\documentclass[preprint,
superscriptaddress,
amsmath,amssymb,aps,showkeys,showpacs,
twoside,final,secnumarabic,
nofootinbib]{revtex4-2}

\usepackage[paperwidth=205mm,paperheight=290mm,top=17mm,bottom=25mm,
inner=17mm,outer=17mm,
twoside]{geometry}

\usepackage{cmap} 
\usepackage[T1,T2A]{fontenc}
\usepackage[utf8]{inputenc}
\usepackage[russian,english]{babel}
\usepackage{color}
\usepackage{graphicx}
\usepackage{dcolumn}
\usepackage{bm} 
\usepackage[unicode=true,colorlinks=true,linkcolor=magenta, urlcolor=blue, citecolor = blue,breaklinks]{hyperref}
\usepackage{multirow}
\usepackage{url}
\usepackage{breakurl}
\DeclareGraphicsExtensions{.eps}

\newcount\issue
\newcount\Vol
\newcount\numb
\usepackage{fancyhdr} 
\def\Vol{\textbf{80}}
\def\numb{x}
\begin{document}

\title{ CONFERENCE SECTION \\[20pt]
Charmonium production in the TMD factorization\\ using the Improved Color Evaporation Model} 

\def\addressa{Samara National Research University, Samara, Russian Federation}
\def\addressb{Joint Institute for Nuclear Research, Dubna, Russian Federation}

\author{\firstname{V.A.}~\surname{Saleev}}
\email[E-mail: ]{saleev.va@ssau.ru}
\affiliation{\addressa}
\affiliation{\addressb}
\author{\firstname{K.K.}~\surname{Shilyaev}}
\email[E-mail: ]{kirill.k.shilyaev@gmail.com}
\affiliation{\addressa}

\received{xx.xx.2025}
\revised{xx.xx.2025}
\accepted{xx.xx.2025}

\begin{abstract}
In this article, the study of unpolarized $J/\psi$ production in proton-proton collisions is presented. The Soft Gluon Resummation approach as a TMD framework was used for description of small-$p_T^{}$ production cross section. The Improved Color Evaporation model was considered as an approach to describe hadronization of produced quarks into charmonium state. We find that the experimental data at various center-of-mass energies $\sqrt{s}$ demonstrate the dependence of hadronization parameter $F^{J/\psi}$ on energy. The predictions for $J/\psi$ production in the kinematics of the SPD NICA experiment have been made.
\end{abstract}

\par
\keywords{high energy physics, quantum chromodynamics, parton model, TMD factorisation, soft gluon resummation approach, improved color evaporation model, charmonium, SPD NICA \\[5pt]}

\maketitle
\thispagestyle{fancy}


\section{Introduction}\label{intro}

The investigation of the $J/\psi$ mesons production in $pp$, $pd$ and $dd$ collisions is a part of the future experimental programme of the SPD NICA collaboration~\cite{1}. In order to describe the production of heavy quarkonia. we should use both the relevant factorization approach and the nonperturbative hadronization model. The analysis of charmonium production in hadron collisions allows us to study both sides of this process. The $J/\psi$ production at small transverse momentum, $p_T^{} < 1$ GeV, is described in the framework of Transverse Momentum Dependent (TMD) factorization~\cite{2}, though some specific approaches are necessary to be employed for modelling TMD parton distributions, so we focus on the Soft Gluon Resummation (SGR) approach in the current study~\cite{3,4}. There are two standard frameworks to describe the processes of hadronization: nonrelativistic QCD (NRQCD)~\cite{5} and Improved Color Evaporation model (ICEM)~\cite{6}. Now we use the latter while the NRQCD was considered in our previous works on $J/\psi$ and $\eta_c^{}$ production within the SGR approach~\cite{7, 8}.

In this study, we present results of calculations and comparison with the data of the production cross sections of unpolarized $J/\psi$ mesons in proton-proton collisions at wide range of center-of-mass energies: from $15$ GeV to $13$ TeV. We also made prediction for the production of unpolarized $J/\psi$ for the forthcoming SPD NICA experiment at $\sqrt{s} = 27$ GeV.

\section{\label{sec:level1}Soft Gluon Resummation approach}

The soft-hard factorization of the production cross section at small transverse momenta of the final state $p_T^{} \ll \mu_F^{}$, where $\mu_F^{}$ is a hard scale of the process, is a subject of the TMD factorization approach~\cite{2, 9}. In the case of charmonium production, its mass $M$ is conventionally taken as a hard scale $\mu_F^{}$. In the TMD factorization, the cross section can be presented as a convolution of TMD parton distribution functions (PDF) and the cross section of the hard partonic subprocess. The TMD PDFs describe the distribution of initial partons over their transverse momentum components and over the longitudinal momentum fraction $x$ of the parent protons momenta. Description of scale evolution of the PDFs demands considering multiple interactions of soft gluons in the initial and final states and, therefore, requires resummation of logarithms of the form $ \alpha_s^{n} (\ln( \mu_ {F}^{2} / p_T^{2} ) )^m$ in all orders with respect to the coupling constant $\alpha_s^{}$. Further, we will consider the SGR approach as a framework of the TMD factorization~\cite{3, 4}.

The momenta of the initial partons $q_{1,2}^\mu$ in the TMD factorization are represented in the standard form of the Sudakov decomposition: 
\begin{equation}
q_{1,2}^\mu = x_{1,2}^{} p_{1,2}^\mu + y_{2,1}^{} p_{2,1}^\mu + q_{1,2T}^\mu,
\end{equation}
where $p_{1,2}^{} = \frac{\sqrt{s}} {2} (1,0,0,\pm 1)$ denotes the momenta of the initial protons, $x_{i}^{}$ and $y_{i}^{} = {\bf q}_{iT}^{2}/(sx_{i})$ are the fractions of proton momenta, $q_{iT}^{}$ are transverse momenta of partons ($q_{iT}^{2} = -{\bf q}_{iT}^{2}$). For small transverse momenta, i.e. preserving corrections up to $\mathcal{O}(|{\bf q}_ {T}^{}|/M)$, the momentum fractions $y_{1,2}^{}$ are negligible, and the parton momenta have the form $q_1^\mu \approx x_1^{} p_1^\mu + q_{1T}^\mu$, so the initial partons are on-shell, $q_{1,2}^ {2} \approx 0$.

The TMD factorization theorem~\cite{2} allows us to represent the cross section for charmonium production in the proton-proton collisions as a convolution of TMD PDFs $F(x,{\bf q}_{T},\mu_F^{},\zeta)$ and the cross section $d\hat{\sigma}$ of the hard partonic subprocess:
\begin{equation}
d\sigma = \int dx_1^{} dx_2^{} \, d^2 q_{1T}^{} \,d^2 q_{2T}^{}\, F (x_1^{}, {\bf q}_{1T}^{}, \mu_{F}^{}, \zeta_1^{}) F (x_2^{}, {\bf q}_{2T}^{}, \mu_{F}^{}, \zeta_2^{})\, d\hat{\sigma},
\end{equation}

The dependence of TMD PDFs on the factorization scale $\mu_F^{}$ and rapidity scale $\zeta$ is described by the Collins–Soper and the renormalization group equations. An analytical factorized solution for this system can only be obtained for the Fourier-transformed PDFs. After a two-dimensional Fourier transform and thereby transition from the transverse momentum space ${\bf q}_{T}^{}$ to the impact parameter space~${\bf b}_{T}^{}$, the solution of the system of equations can be obtained in the following form~\cite{10}:
\begin{equation}
\hat{F}(x_1^{}, b_{T}^{}, \mu_{F}^{}, \zeta) \hat{F}(x_2^{}, b_{T}^{}, \mu_{F}^{}, \zeta)  = e^{-S_P^{} (b_{T}^{},\mu_{F}^{}, \mu_{F0}^{}, \zeta, \zeta_{0}^{})} \hat{F}(x_1^{}, b_{T}^{}, \mu_{F0}^{}, \zeta_{0}^{}) \hat{F}(x_2^{}, b_{T}^{}, \mu_{F0}^{}, \zeta_{0}^{}),
\end{equation}
where $S_P^{} (b_{T}^{},\mu_{F}^{}, \mu_{F0}^{}, \zeta, \zeta_{0}^{})$ denotes the Sudakov factor, which evolves TMD PDFs from initial scales ($\mu_{F0}^{}$, $\zeta_0$) to final scales ($\mu_{F}^{}$, $\zeta$). For proton-proton collisions in the leading logarithmic (LL) approximation and the leading order with respect to the coupling constant $\alpha_s^{}$ (LO), the Sudakov factor has the form~\cite{3, 11}:
\begin{widetext}
\begin{equation}
S_P^{} (\mu_{F}^{}, \mu_b, b_T^{}) = \frac{C_A^{}}{\pi} \int\limits_{\mu_b^2}^{\mu_{F}^{2}} \frac{d\mu'^{2}}{\mu'^{2}} \alpha_s (\mu') \left[ \ln \frac{\mu_{F}^{2}}{\mu'^2} - \left( \frac{11-2N_f/C_A}{6} + \frac{1}{2} \right) \right] + \mathcal{O}(\alpha_s),
\end{equation}
\end{widetext}
where $N_f$ is the number of quark flavours, $C_A = N_c = 3$, the initial scale is $\mu_b^{} = \mu_{F0}^{} = \sqrt{\zeta_0^{}} \sim b_0^{}/b_T^{}$. In the one-loop approximation for the coupling constant $\alpha_s^{}$, the integral can be explicitly evaluated and an expression for the Sudakov factor $S_P^{}$ in LL-LO can be obtained. However, this expression is only valid in the range of the impact parameter $b_0^{}/ \mu_{F}^{} \leqslant b_T^{} \leqslant b_{T,\, \textnormal{\scriptsize max}}^{}$, where $b_0^{}=2e^{-\gamma}$, $\gamma$ is the Euler-Mascheroni constant, $b_{T,\, \text{max}}$ is some maximum value of the aiming parameter that cannot be determined from first principles. This upper limit is realized by cutting off the impact parameter~\cite{12}: $b_T^{*} (b_T^{}) = b_T^{}/\sqrt{1+(b_T^{}/b_{T,\,\text{max}}^{})^2}$ with a maximum value of $b_{T,\,\text{max}} = 1.5$ GeV$^{-1}$. The lower limit of the range is given by the expression $\mu_b'^{} = b_0^{}/(b_T^{} + b_0^{}/\mu_{F}^{})$.

Suppression of the SGR PDFs at large $b_T^{}$ is also guaranteed by the nonperturbative Sudakov factor $\tilde{S}_{NP}^{}$, the expression for which is taken in the form of Gaussian ansatz. In our work, we use the following parametrization for initial quarks: 
\begin{equation}
\tilde{S}_{NP}^{} (x, b_T^{}, \mu_{F}^{}) = \frac{1} {2} \left[ g_1^{} \ln \frac{\mu_{F}^{}}{2Q_{NP}^{}} + g_2^{} \left( 1 + 2 g_3^{} \ln \frac{10 x x_0^{}}{x_0^{} + x} \right) \right] b_T^2,
\end{equation}
with the following model parameter values obtained  from experimental SIDIS data: $g_1^{} = 0.184$ GeV$^2$, $g_2^{} = 0.201$ GeV$^2$, $g_3^{} = -0.129$, $x_0^{} = 0.009$, $Q_{NP}^{} = 1.6$ GeV~\cite{13}. To apply the factor to initial gluons, it requires color factor change with $C_A^{}/C_F^{}$, where $C_A^{} = N_c^{} = 3$, $C_F^{} = (N_c^2 - 1)/(2N_c^{}) = 4/3$, these are the eigenvalues of the Casimir operator of the adjoint and fundamental representations of the SU($3$) group, respectively~\cite{11}. Because SGR PDF for each of the protons are included in the cross section, the nonperturbative Sudakov factor should be written as follows: $S_{NP}^{} (x_1^{}, x_2^{}, b_T^{}, \mu_{F}^{}) = \tilde{S}_{NP}^{} (x_1^{}, b_T^{}, \mu_{F}^{}) + \tilde{S}_{NP}^{} (x_2^{}, b_T^{}, \mu_{F}^{})$.

On a small initial scale $\mu_b'^{}$ the dependence of Fourier transformed SGR PDFs is expressed via collinear PDFs, which is in the LO approximation as follows:
\begin{equation}
\hat{F}(x, \mu'^{}_{b}, b_T^{}) = f (x, \mu'^{}_{b}) + \mathcal{O} (\alpha_s) + \mathcal{O} (b_T^{} \Lambda_{\text{QCD}}).
\end{equation}

Thus, the expression for the cross section of quark-antiquark pair production at $p_T^{c\bar{c}} \ll M_{c\bar{c}}^{}$ can be written as
\begin{equation}
\frac{d\sigma}{dp_{T}^{c\bar{c}} \, dy \, dM_{c\bar{c}}^2} = \frac{p_T^{c\bar{c}}}{s} \int  db_T^{} \, b_T^{} \, J_0(b_T^{}p_T^{c\bar{c}} ) \, e^{-S_P^{}} \, e^{-S_{NP}^{}} \, \hat{F}_1^{} \, \hat{F}_2^{} \, \hat{\sigma}(M_{c\bar{c}}^2),
\label{eq:sgr}
\end{equation}
where $S_P^{} \equiv S_P^{} (\mu_{F}^{},  \mu'_{b^{*}}, b_T^{*})$ and $S_{NP}^{} \equiv S_{NP}^{}(x_1^{}, x_2^{}, b_T^{}, \mu_{F}^{})$ are Sudakov factors, $\hat{F}_1^{} \equiv \hat{F} (x_1^{},  \mu'^{}_{b^{*}}, b_T^{*})$ and~$\hat{F}_2^{} \equiv \hat{F} (x_2^{},  \mu'^{}_{b^{*}}, b_T^{*})$ are PDFs, $J_0$ is a Bessel function of the first kind of zeroth order, $p_{T}^{\,c\bar{c}}$, $y$ and $M_{c\bar{c}}^{}$ --- transverse momentum, rapidity and invariant mass of the final $c\bar{c}$ pair, $\hat{\sigma} (M_{c\bar{c}}^2)$ is the total cross section for production of the $c\bar{c}$ pair with invariant mass $M_{c\bar{c}}^{}$. The total cross sections for $c\bar{c}$ pair production in gluon-gluon fusion and quark-antiquark annihilation subprocesses are written explicitly as:
\begin{equation}
\hat{\sigma}_{gg \rightarrow c\bar{c}} (\hat{s}) = \frac{\pi\alpha_s^2}{3\hat{s}} \left[ \left(1 + w + \frac{w^2}{16} \right) \ln \left( \frac{1+\sqrt{1-w}}{1-\sqrt{1-w}} \right) - \left( \frac{7}{4}  + \frac{31 w}{16}\right) \sqrt{1-w} \right],
\end{equation}
\begin{equation}
\hat{\sigma}_{q\bar{q} \rightarrow c\bar{c}}(\hat{s}) = \frac{4 \pi \alpha_s^2}{27 \hat{s}} \left( w + 2 \right) \sqrt{1-w},
\end{equation}
where $w = 4 m_c^2/\hat{s}$, $m_c^{}$ is a mass of $c$-quark, $\hat{s} = (q_1^{} + q_2^{})^2 = (p^{c} + p^{\bar{c}})^2$.

\section{\label{sec:level1}Improved Color Evaporation model}

To describe hadronization of the produced $c\bar{c}$-quark pair into observable state of charmonium, we use the ICEM~\cite{6}. In this framework, it is assumed that a pair of heavy quarks is produced with the invariant mass $M_{c\bar{c}}$, followed by soft gluon emissions and interactions of the final quark-antiquark pair with the other color sources of the process, in this way the observable quarkonium state with mass $M_{J/\psi}$ is formed.
 
The cross section for the $J/\psi$ production in the ICEM is represented by averaging the cross section of $c\bar{c}$-pair production over the invariant mass $M_{c\bar{c}}^{}$ in the following range: $M_{J/\psi}^{} \leqslant M_{c\bar{c}}^{} \leqslant 2M_{D}^{}$, where $M_{D}^{}$ is the mass of the lightest $D$-meson. Also, in the ICEM, the cross section should be multiplied by a nonperturbative phenomenological factor $F^{J/\psi}$, which can be interpreted as the probability of heavy quark-antiquark pair transition to $J/\psi$ state. The formation of $J/\psi$ from the $c\bar{c}$-pair also requires the following 4-momentum shift: $p_{\mu}^{J/\psi} = p_{\mu}^{c\bar{c}} \cdot M_{J/\psi}^{} / M_{c\bar{c}}^{}$.

As a result, the expression for the differential cross section of $J/\psi$ production in the ICEM is written as follows:
\begin{widetext}
\begin{equation}
\frac{d\sigma}{d^2p_{T}^{J/\psi} \, dy} = F^{J/\psi} \int\limits_{M_{J/\psi}^{2}}^{4M_ {D}^{2}} dM_{c\bar{c}}^{2} \int d^2 p_{T}^{c\bar{c}} ~ \delta^{(2)} \left( {\bf p}_{T}^{\,J/\psi} - \frac{M_{J/\psi}^{}}{M_{c\bar{c}^{}}} {\bf p}_{T}^{\,c\bar{c}} \right) \, \frac{d\sigma}{d^2p_{T}^{c\bar{c}} \, dy \, dM_{c\bar{c}}^2},
\label{eq:icem}
\end{equation}
\end{widetext}
where the $\delta$-function implements the momentum shift, and $d\sigma/d^2p_{T}^{c\bar{c}} \, dy \, dM_{c\bar{c}}^2$ is the differential cross section of $c\bar{c}$-pair production. Then, substituting the expression (\ref{eq:sgr}) into (\ref{eq:icem}), we obtain the final expression for the differential cross section in the SGR and ICEM:
\begin{widetext}
\begin{equation}
\frac{d\sigma}{dp_{T}^{J/\psi} \, dy} = F^{J/\psi} \,\frac{p_T^{\,J/\psi}}{s} \int\limits_{M_{J/\psi}^{2}}^{4M_{D}^{2}} dM_{c\bar{c}}^{2} \, \frac{M^2_{c\bar{c}}}{M^2_{J/\psi}} \int\limits_0^{\infty} db_T^{} \, b_T^{} \, J_0 \left( \frac{M_{c\bar{c}}}{M_{J/\psi}} p_T^{J/\psi}  b_T^{} \right) e^{-S_P^{}} \, e^{-S_{NP}^{}} \, \hat{F}_1^{} \, \hat{F}_2^{} \, \hat{\sigma}(M_{c\bar{c}}^2)
\end{equation}
\end{widetext}
with the same notations as those were used in (\ref{eq:sgr}).

\section{\label{sec:level1}Results}

In our numerical calculations, the collinear PDFs in the leading order with respect to $\alpha_s^{}$ were taken as tabulated MSTW2008LO distributions ~\cite{14}. The following quantities were used in the calculations~\cite{15}: $M_{J/\psi}^{} = 3.096$ GeV and $M_{D}^{} = 1.87$ GeV, branching ratios $\text {Br}(J/\psi \rightarrow e^{+}e^{-}) = 0.05971$, $\text{Br}(J/\psi \rightarrow \mu^{+}\mu^{-}) = 0.05961$. We took the invariant mass of the quark-antiquark pair $M_{J/\psi}^{}$ as the factorization scale $\mu_{F}^{}$ and renormalization scale $\mu_{R}^{}$.

We describe experimental data for the wide range of the centre-of-mass energies: from $15$ GeV to $13$ TeV. Within the framework of the ICEM formalism, we make predictions for prompt $J/\psi$ production which includes contributions from feed-down decays of excited charmoinum states. As it was shown in Ref.~\cite{16}, the factor $F^{J/\psi}$ depend on $\sqrt{s}$, so in the current study work we extract the values of $F^{J/\psi}$ for each $\sqrt{s}$. 

\begin{figure*}
\includegraphics[scale=0.3]{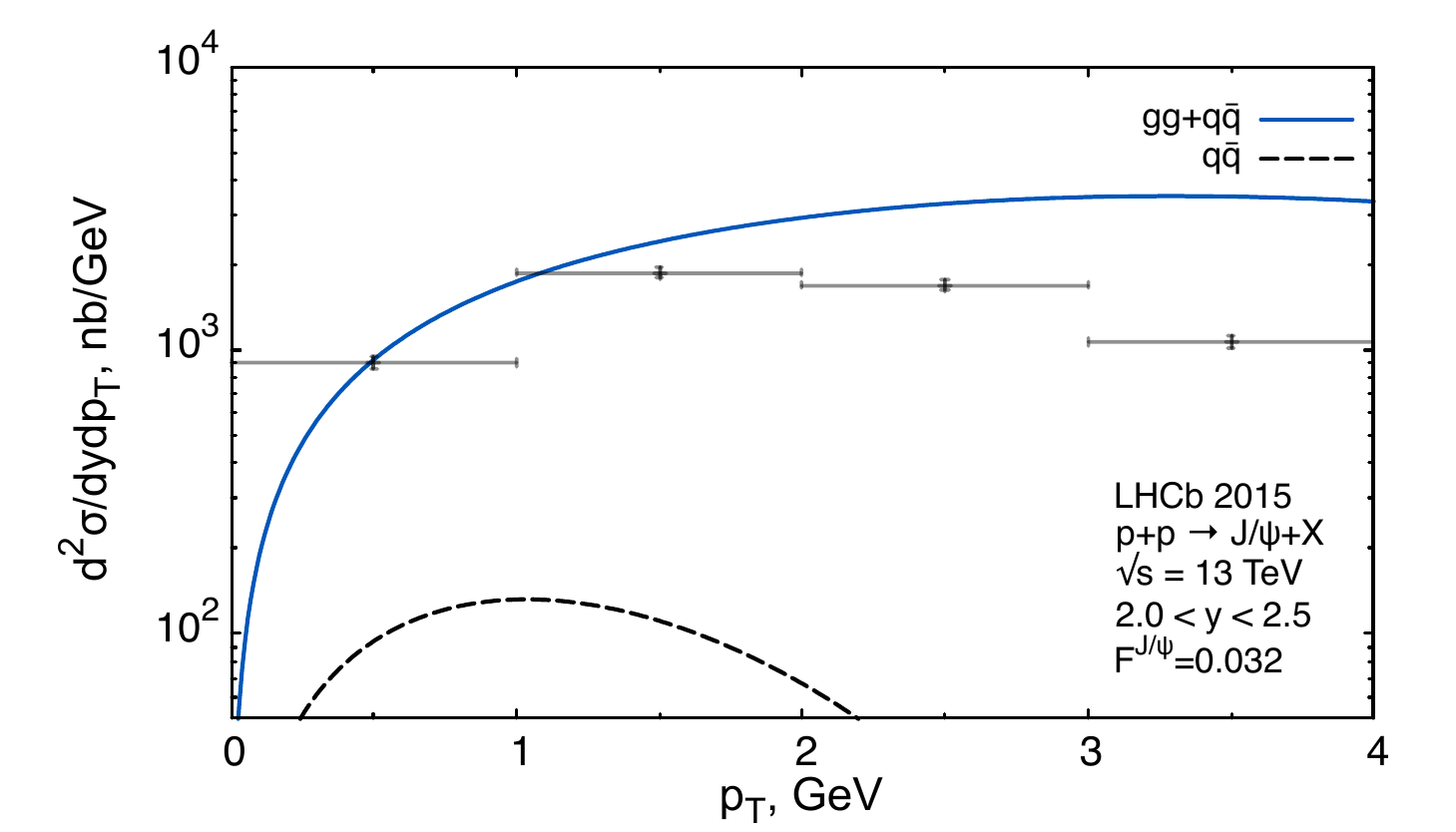}
\includegraphics[scale=0.3]{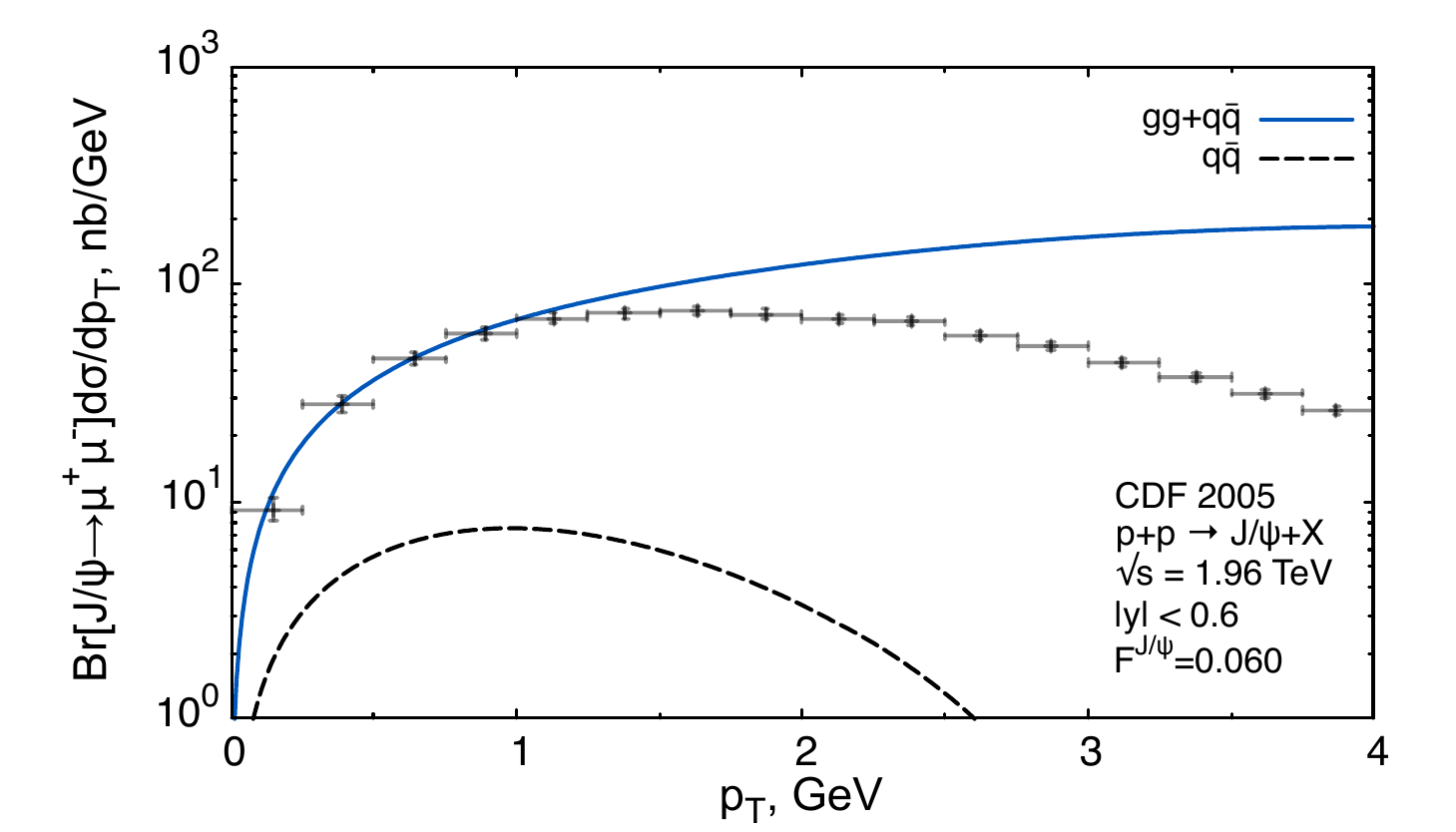}
\caption{\label{fig:1}Diﬀerential cross section of $J/\psi$ production with the experimental data of LHCb Collaboration~\cite{17} (on the left) and CDF Collaboration~\cite{18} (on the right).}
\end{figure*}

\begin{figure*}
\includegraphics[scale=0.3]{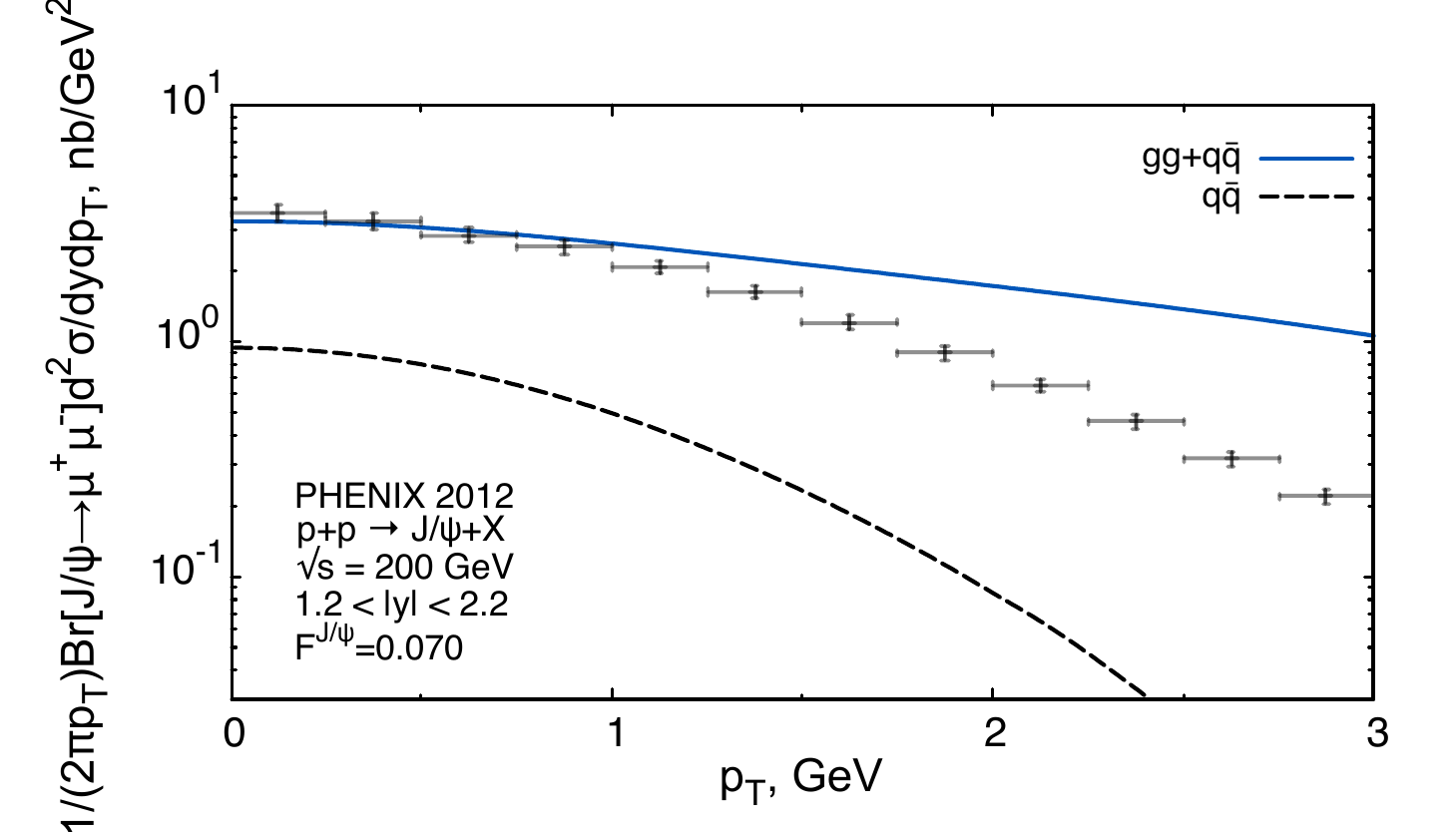}
\includegraphics[scale=0.3]{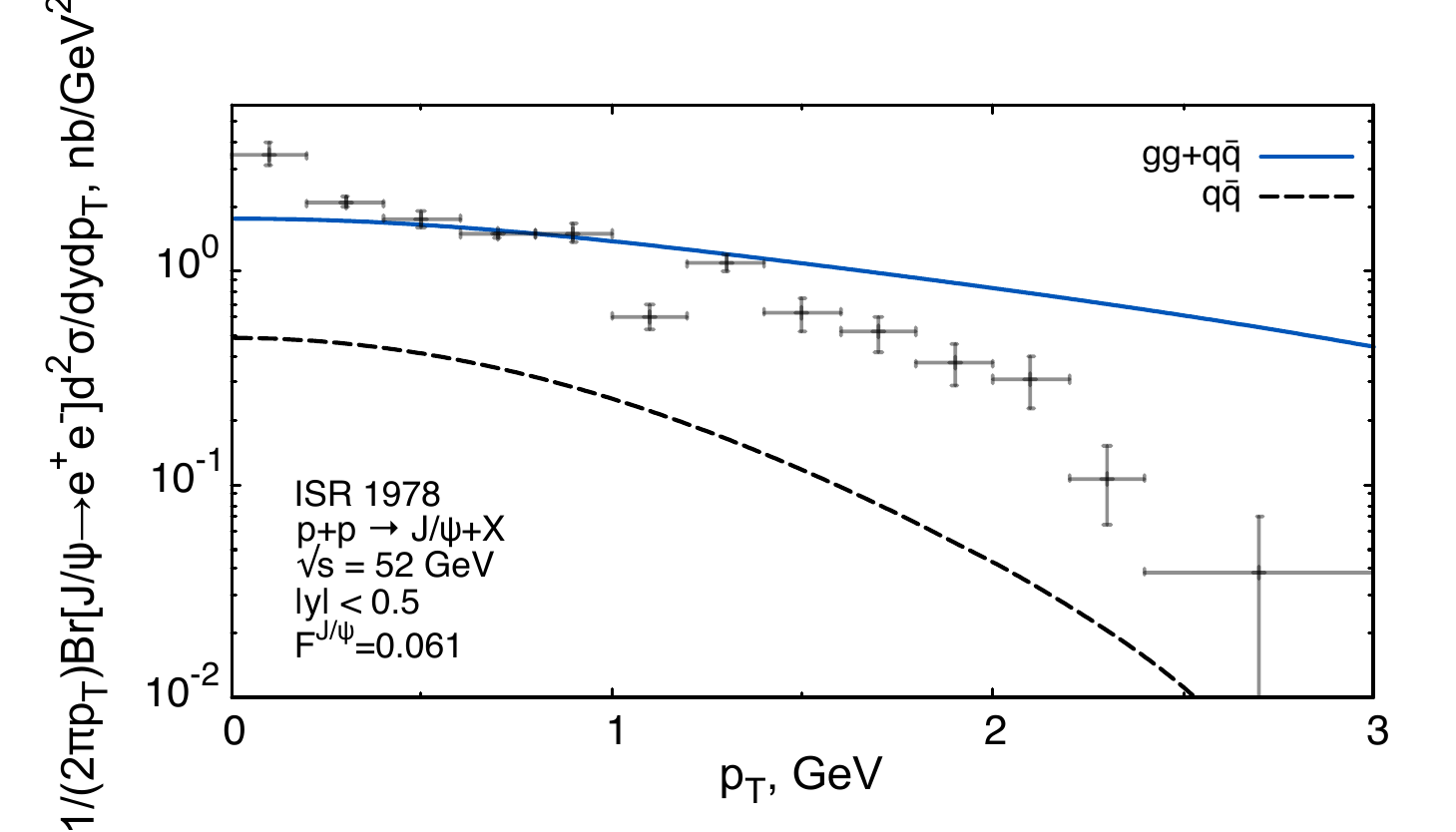}
\caption{\label{fig:2}Diﬀerential cross section of $J/\psi$ production with the experimental data of PHENIX Collaboration~\cite{19} (on the left) and from ISR CERN~\cite{20} (on the right).}
\end{figure*}

\begin{figure*}
\includegraphics[scale=0.3]{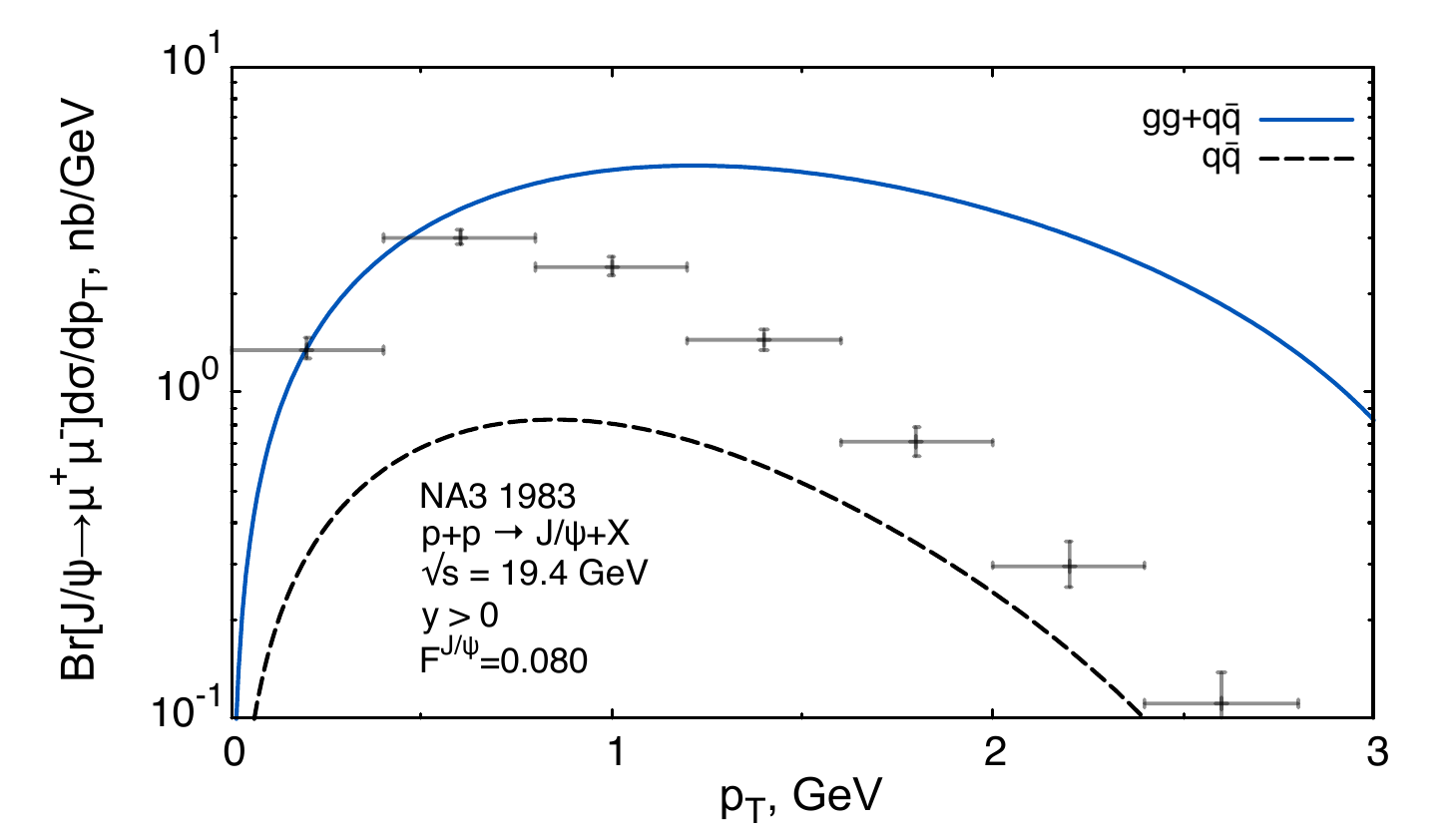}
\includegraphics[scale=0.3]{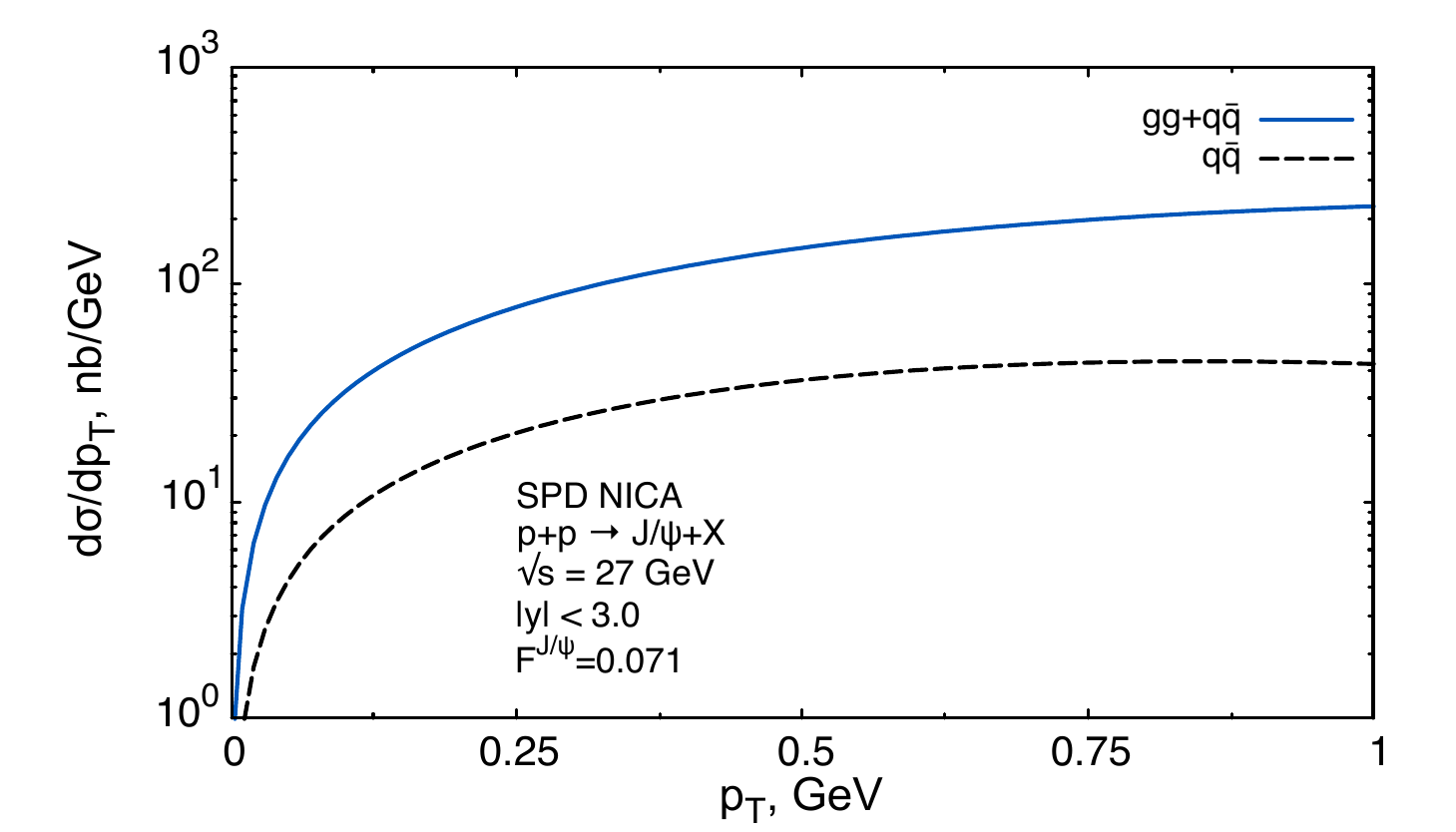}
\caption{\label{fig:3}Diﬀerential cross section of $J/\psi$ production with the experimental data of NA3 Collaboration~\cite{21} (on the left) and prediction for SPD NICA (on the right).}
\end{figure*}

In the Figs.~\ref{fig:1}-\ref{fig:3}, the results of our calculations in the SGR approach using the ICEM for various $\sqrt{s}$ values are demonstrated. The $F^{J/\psi}$ values fitted to experimental data in the charmonium transverse momentum range $p_T^{} < 1$ GeV are also shown in the figures. These plots show the contributions of quark-antiquark annihilation subprocesses separately. Although gluon-gluon fusion processes dominate for all energies, the fraction of quark-antiquark annihilation processes increases when $\sqrt{s}$ decreases. The contribution of quark-antiquark annihilation is shown with dashed lines on the plots, the sum of both hard processes is shown with solid lines.

In addition, Fig.~\ref{fig:3} shows predictions for differential cross sections of $J/\psi$ production in the kinematics of the future SPD NICA experiment. In order to estimate $F^{J/\psi}$ for $\sqrt{s} = 27$ GeV, we take the arithmetic mean of the values of this parameter for $\sqrt{s} = 19.4$ GeV and $\sqrt{s} = 52$~GeV, i.e. for SPD NICA $F^{J/\psi} = 0.071$.\\

\section*{FUNDING}

The work was supported by the Foundation for the Advancement of Theoretical Physics and Mathematics BASIS, grant No. 24-1-1-16-5, and by the grant of the Ministry of Science and Higher Education of Russian Federation, No. FSSS-2025-0003.

\section*{CONFLICT OF INTEREST}

The authors of this work declare that they have no conflicts of interest. 



\begin{thebibliography}{}

\bibitem{1}
A. Arbuzov, A. Bacchetta, M. Butenschoen, et al., Prog. Part. Nucl. Phys \textbf{119}, 103858 (2021). https://doi.org/10.1016/j.ppnp.2021.103858.

\bibitem{2}
J. Collins, in \emph{Foundations of Perturbative QCD}, (Cambridge University Press, 2011), p. 624. https://doi.org/10.1017/9781009401845.

\bibitem{3}
D. Boer and W. J. den Dunnen, Nucl. Phys. B \textbf{886} (2014), 421. https://doi.org/10.1016/j.nuclphysb.2014.07.006.

\bibitem{4}
P. Sun, B. W. Xiao, and F. Yuan, Phys. Rev. D \textbf{84}, 094005 (2011). https://doi.org/10.1103/PhysRevD.84.094005.

\bibitem{5}
G. T. Bodwin, E. Braaten and G. P. Lepage, Phys. Rev. D \textbf{51}, 1125 (1995). https://doi.org/10.1103/PhysRevD.55.5853.

\bibitem{6}
Y. Q. Ma and R. Vogt, Phys. Rev. D \textbf{94}, 11, 114029 (2016). https://doi.org/10.1103/PhysRevD.94.114029.

\bibitem{7}
V. A. Saleev, K. K. Shilyaev, [arXiv:2502.16461 [hep-ph]].

\bibitem{8}
V. A. Saleev and K. K. Shilyaev, Phys. Atom. Nucl. \textbf{88}, 2, 338 (2025). https://doi.org/10.1134/S1063778825700309

\bibitem{9}
R. Boussarie, M. Burkardt, M. Constantinou, et al. [arXiv:2304.03302 [hep-ph]].

\bibitem{10}
J. C. Collins and D. E. Soper, Nucl. Phys. B \textbf{193}, 381 (1981). https://doi.org/10.1016/0550-3213(81)90339-4.

\bibitem{11}
J. Bor and D. Boer, Phys. Rev. D \textbf{106}, 1, 014030 (2022). https://doi.org/10.1103/PhysRevD.106.014030.

\bibitem{12}
J. C. Collins, D. E. Soper, and G. F. Sterman, Nucl. Phys. B \textbf{250}, 199 (1985). https://doi.org/10.1016/0550-3213(85)90479-1.

\bibitem{13}
S. M. Aybat and T. C. Rogers, Phys. Rev. D \textbf{83}, 114042 (2011). https://doi.org/10.1103/PhysRevD.83.114042.

\bibitem{14}
A. D. Martin, W. J. Stirling, R. S. Thorne, and G. Watt, Eur. Phys. J. C \textbf{63}, 189 (2009).
https://doi.org/10.1140/epjc/s10052-009-1072-5.

\bibitem{15}
P. A. Zyla, R. M. Barnett, J. Beringer, O. Dahl, et al., [Particle Data Group], PTEP \textbf{2020}, 8, 083C01 (2020). https://doi.org/10.1093/ptep/ptaa104.

\bibitem{16}
A. A. Chernyshev and V. A. Saleev, Phys. Rev. D \textbf{106},11,114006 (2022). https://doi.org/10.1103/PhysRevD.106.114006

\bibitem{17}
R. Aaij, B. Adeva, M. Adinolfi, A. Affolder, Z. Ajaltouni, et al., JHEP \textbf{10}, 172 (2015). https://doi.org/10.1007/JHEP10(2015)172.

\bibitem{18}
D. Acosta, J. Adelman, T. Affolder, T. Akimoto, M. G. Albrow, et al., Phys. Rev. D \textbf{71}, 032001 (2005). https://doi.org/10.1103/PhysRevD.71.032001

\bibitem{19}
A. Adare, S. Afanasiev, C. Aidala, N. N. Ajitanand, Y. Akiba, et al., Phys. Rev. D \textbf{85}, 092004 (2012). https://doi.org/10.1103/PhysRevD.85.092004.

\bibitem{20}
A. G. Clark, P. Darriulat, K. Eggert, V. Hungerbuhler, H. R. Renshall, et al., Nucl. Phys. B \textbf{142}, 29 (1978). https://doi.org/10.1016/0550-3213(78)90400-5.

\bibitem{21}
J. Badier, J. Boucrot, J. Bourotte, G. Burgun, O. Callot, et al., Z. Phys. C \textbf{20}, 101 (1983). https://doi.org/10.1007/BF01573213.

\end{thebibliography}
\end{document}